\pdfoutput=1

\documentclass{sig-alternate}
\usepackage{graphicx}
\usepackage{authordate1_4}
\usepackage{url}
\usepackage{balance}

\begin{document}
%

\conferenceinfo{LAK '14,}{March 24 -- 28 2014, Indianapolis, IN, USA}
\CopyrightYear{2014}
\crdata{978-1-4503-2664-3/14/03}

\title{Modelling student online behaviour in a virtual learning environment}
%
%
%
%
%

\numberofauthors{4} 
%
\author{
%
%
Martin Hlosta\textsuperscript{\textdagger} \quad Drahomira Herrmannova\textsuperscript{\textdagger} \quad Lucie Vachova\textsuperscript{\textdagger\textdagger} \\ Jakub Kuzilek\textsuperscript{\textdagger} \quad Zdenek Zdrahal\textsuperscript{\textdagger} \quad Annika Wolff\textsuperscript{\textdagger} \\ \\
    \begin{tabular}[t]{@{}c@{}}
    \affaddr{Knowledge Media Institute, The Open University\textsuperscript{\textdagger}} \\
    \affaddr{Walton Hall} \\
    \affaddr{Milton Keynes, MK7 6AA} \\
    \email{\{martin.hlosta; d.herrmannova; jakub.kuzilek;} \\
    \email{z.zdrahal; annika.wolff\}@open.ac.uk}
    \end{tabular}\nobreak\qquad
    \begin{tabular}[t]{@{}c@{}}
    \affaddr{University of Economics, Prague\textsuperscript{\textdagger\textdagger}\titlenote{This work was carried out at the Knowledge Media Institute.}} \\
    \affaddr{Department of Exact Methods, Faculty of Management} \\
    \affaddr{Jarosovska 1117/II} \\
    \affaddr{Jindrichuv Hradec, 377 01} \\
    \email{vachova@fm.vse.cz}
  \end{tabular}
}
\date{\today}

\maketitle
\begin{abstract}
In recent years, distance education has enjoyed a major boom. Much work at The Open University (OU) has focused on improving retention rates in these modules by providing timely support to students who are at risk of failing the module. In this paper we explore methods for analysing student activity in online virtual learning environment (VLE) -- \emph{General Unary Hypotheses Automaton (GUHA)} and \emph{Markov chain-based analysis} -- and we explain how this analysis can be relevant for module tutors and other student support staff. We show that both methods are a valid approach to modelling student activities. An advantage of the Markov chain-based approach is in its graphical output and in the possibility to model time dependencies of the student activities.
\end{abstract}

\category{\\ H.2.8}{Database Applications}{Data Mining}




\section{Introduction}

The recent years have seen a massive growth of different possibilities of online education, such as the well known massive open online courses (MOOCs) \cite{Cormier:2008:MOOC}. The concept of distance education is however not new. The Open University is an institution with over forty years of experience with distance education, historically based on off-line materials and nowadays making an increasing use of the Internet. The great advantage of the online courses is in the fact they are accessible to virtually anyone with Internet access.

The other side of the coin is that the retention rates in these courses are often low. \cite{Koller:2013:MOOC_Retention} mention, that an average retention rate of a Coursera\footnote{\url{https://www.coursera.org/}, a well known and one of the biggest platforms providing open online courses.} course is around 5\%. The situation at traditional universities as well as at The Open University is significantly better, however, there is still a room for improvement.

There might be many reasons for the low retention rates, from the fact that the online courses are often offered to anybody interested to the fact that the performance of each student depends almost exclusively on how much are they willing to study on their own at home. Our work at The Open University aims at analysing students' activities in the online courses in order to gain insight into their behavioural patterns, which can be utilised for building prediction models.

\subsection{Problem Description}

The Open University\footnote{\url{http://www.open.ac.uk/}} is the biggest university in the Unit\-ed Kingdom, offering several hundred distance learning modules, which can be studied both as standalone modules or as part of a university degree. Anybody can sign up for a module provided by The OU, without any previous education whatsoever. The students receive their study material and submit their assignments through an online virtual learning environment.

Students participating in a module are generally split into smaller study groups of no more than few tens of students, typically according to their geographic location. Each group has an assigned tutor. The tutors grade the students' assignments and exams, answer their questions in the online forums, provide general advice and guidance, etc.

In order to support the students who are at risk of failing the module The OU also implements various interventions (such as phone calls from a specialised student support teams) during the course of the module. Because the number of students studying each module can reach several thousand (the modules used in our analysis have enrolment of around two thousand students) and the resources available for the interventions are limited, the interventions have to be carefully planned. Therefore, an important question one might ask is how to identify students at risk of failing the module so that the intervention is meaningful and efficient.

Improving student retention through these focused interventions and helping the tutors to focus on the students, who require help, provides many benefits, from improved student satisfaction to financial savings for the university. In~order to support the identification of students who are currently at risk, we utilise several statistical and machine learning methods. The available data contain both the information about the students' activity in the VLE as well as their demographic information. However, for modelling student behaviour, only the data from the VLE was used. A more detailed description of our data set can be found in Section \ref{sec:data_specification}.


\section{Previous Work}

The current work builds on previous research done at The Open University. The initial experiments with machine learning techniques were using the VLE and assessment data \cite{Wolff:2012}. One of the main findings of this research was that decision trees generally outperform the other methods \cite{Wolff:2012}. This research also included creation of a dashboard providing the university staff with real-time information about student performance.

Additional methods were tested in \cite{Wolff:2013} and in \cite{Wolff:2014}. In the latter work, demographic data were added to the predictions, however this research did not confirm that this data provide a significant increase in performance. 


\section{Data Specification}
\label{sec:data_specification}


The analyses we performed were done using real data of several modules from The Open University. We examined a number of subsequent presentations of each module.

The available data contain two types of information:

\begin{itemize}
    \item Information about the results of student assignments (TMAs -- tutor marked assignments). There are several assignments in each module, typically between five and seven. Generally, the module is ended by a final exam.
    \item Data about student activity from the virtual learning environment (VLE).
\end{itemize}

The VLE data are aggregated by days and content type (e.g. forum, wiki, resource, ...). This means that for each day we know how many times did the student interact with given content type. For our analysis we summarise the data by weeks and content types. Summarising the data by weeks seemed to be reasonable, it simplifies our analyses without loosing too much detail. The features generated by the summarisation and used in the methods are:

\begin{itemize}
  \item click counts aggregated by week,
  \item click counts aggregated by week and content type,
  \item binary flags indicating whether student was active in the VLE and in various content types.
\end{itemize}


\section{Methods}

For analysis of student behaviour in the virtual learning environments, we have used two different approaches -- GUHA \cite{Hajek:1966:GUHA} and modelling based on Markov chains \cite{norris1998markov}.


\subsection{Activity types analysis}

As mentioned in Section \ref{sec:data_specification}, the VLE data contain information about the type of content the student accessed. The content type can be for example \textit{forum}, \textit{wiki}, \textit{resource}, \textit{quiz}, etc. In total there are 11 different content types. Using the binary flags, indicating whether student was active in given week and content type, we utilised Bayes Theorem \cite{bishop2006pattern} for determining the probability that the student will fail to complete the module. Moreover, we analysed each of the content types in terms of mean number of students succeeding in the module based on activity or inactivity in the given content type. Based on this investigation we have selected a set of content types which significantly influence students' performance, these were then used in the further analyses.


\subsection{GUHA}

\emph{General Unary Hypotheses Automaton (GUHA)}, originally published in \cite{Hajek:1966:GUHA}, is one of the oldest data mining methods for automatic discovery of new interesting hypotheses from the data. To achieve this goal, GUHA uses various different procedures (ASSOC, IMPL, CORREL). The choice of the procedure to use depends mainly on the user needs and his experience. We have selected the \emph{ASSOC} procedure \cite{Rauch:2001:4ft}, which allows to discover interesting associations between attributes in the data. The interestingness of the association is mostly based on their co-occurrence. The ASSOC procedure allows to limit the resulting rules by specifying constraints on the attributes. This property is important for our field of interest.

For our research, we used the ASSOC procedure that is implemented in the 4ft-Miner module  within the Lisp-Miner software tool \footnote{LISp-Miner \url{lispminer.vse.cz/} -- software tool for implementation of the GUHA method.}. The specification of the constraints enabled us to restrict the rules only to those, which cover students that fail or succeed in the TMA. We used three basic types of features for the analysis introduced in Section \ref{sec:data_specification}.

The search space for both types of binary flags was reasonable to perform analysis. For weekly aggregated counts, it was necessary to reduce the search space via interval discretisation. For this purpose, we utilized LISp-miner and unsupervised equal frequency discretisation \cite{Wong:1987:discretization}.

\begin{figure}
  \centering
  \includegraphics[width=\columnwidth]{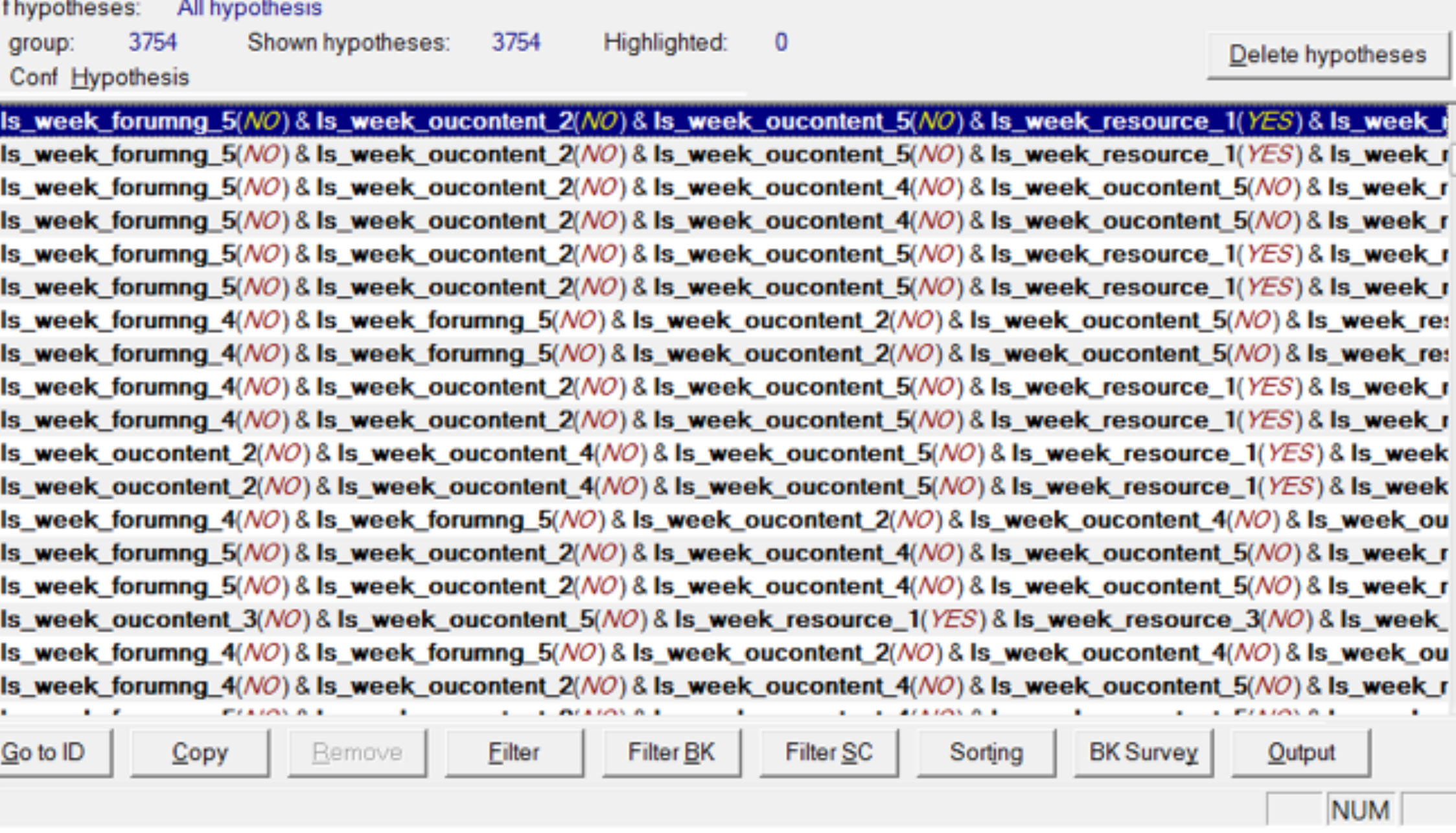}
  \caption{Screenshot with discovered rules from LISp-miner.}
  \label{fig:guha_flags}
\end{figure}

GUHA method produce large set of hypotheses. The example of such results with the categorized binary flags are depicted in the Figure \ref{fig:guha_flags}. Unfortunately the information contained in the output is  difficult to interpret. Moreover, the information of the time dimension is lost and this is even worse when using various content types. For us, this was the motivation to look for another modelling method.

\subsection{Markov chain-based analysis of student activity}

In this part of analysis we examined the differences in intensity of student activity between students who were successful in the first TMA (TMA 1) and those who did not submit TMA 1. Students who failed in TMA 1 are not included in this analysis, due to the fact that they represent only a small portion of students who submit TMA 1. In this stage of research we analysed only students who at least once had zero VLE activity in one of the weeks under consideration. In VLE passive students represent the specific group important while looking for potentially at-risk students. On this group we have studied different scenarios - the list of them is displayed in Table \ref{tab:markov}. Moreover, the Table \ref{tab:markov} shows percentage of students who behaved according to given Scenario and were successful in TMA 1, in comparison to those who did not submit TMA 1, as well. Numbers in column Scenario represents particular weeks of course.

\begin{table}
    \begin{tabular}{|l|p{40mm}|r|r|}
    \hline
    ~   & Scenario                                                     & TMA 1    & TMA 1 \\
    ~   & ~                                                            & not submit &  success \\ \hline \hline
    1.  & zero in any of 0 - 4                                   & 51.6                                                       & 42.6                                                    \\
    \textbf{2.}  & \textbf{zero only in 1 - 4}                                     & \textbf{95.7}                                                       & \textbf{4.3}                                                     \\
    \textbf{3.}  & \textbf{zero only in 2 - 4}                                     & \textbf{92.3}                                                       & \textbf{7.7}                                                     \\
    \textbf{4.}  & \textbf{zero only in 3 - 4}                                     & \textbf{100.0}                                                      & \textbf{0.0}                                                     \\
    5.  & zero only in 4                                          & 54.3                                                       & 45.7                                                    \\
    6.  & zero only in 0                                          & 15.4                                                       & 71.8                                                    \\
    \textbf{7.}  & \textbf{zero only in 0 - 1}                                      & \textbf{6.7}                                                        & \textbf{86.7}                                                    \\
    8.  & zero only in 0 - 2                                      & 15.4                                                       & 69.2                                                    \\
    9.  & zero only in 0 - 3                                      & 57.1                                                       & 42.9                                                    \\
    10. & zero in at least one of 0 - 3, non-zero in  4      & 18.2                                                       & 71.6                                                    \\
    11. & zero in at least one of 0 - 2, non-zero in  3 - 4 & 14.6                                                       & 75.1                                                    \\
    12. & zero in at least one of 0 - 1, non-zero in  2 - 4 & 9.3                                                        & 80.4                                                    \\ \hline
    \end{tabular}
    \caption{Summary of examined situations for students behaviour evaluation (data in \% of total number of course students)}
    \label{tab:markov}
\end{table}

Based on the data in Table \ref{tab:markov}, we can identify behaviour of at-risk students. This is evident especially in scenarios 2, 3, 4 and 7. This shows us that students who tend to reach zero VLE activity in later weeks are more probable not tu submit (scenarios 2, 3, 4). On the other hand those who have zero VLE activity in earlier weeks and later start to show interest represented by VLE activity, raise their chance to success in TMA 1 (scenarios 6, 7, 8).

\begin{figure}
  \centering
  \includegraphics[width=\columnwidth]{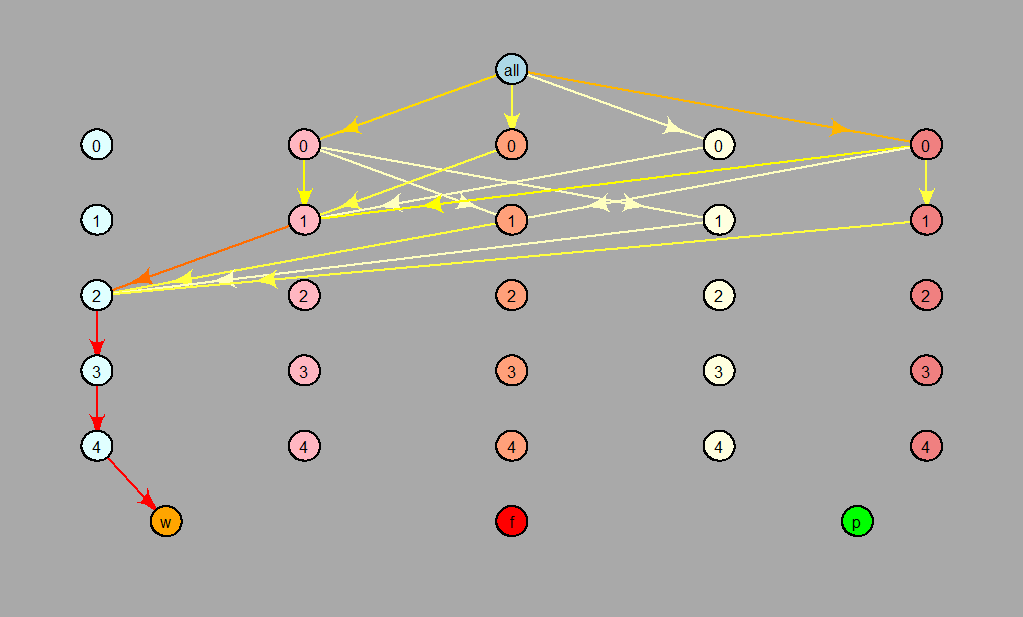}
  \caption{Representation of students behaviour in scenario 3 (TMA 1 was not submitted)}
  \label{fig:markov_xx}
\end{figure}

\begin{figure}
  \centering
  \includegraphics[width=\columnwidth]{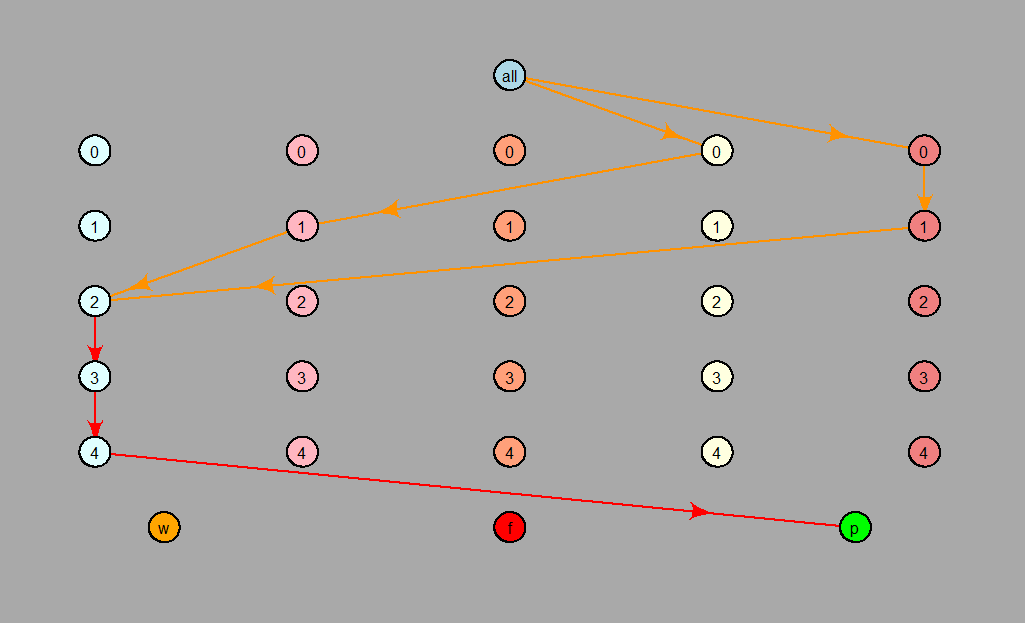}
  \caption{Representation of students behaviour in scenario 3 (TMA 1 was passed)}
  \label{fig:markov_xy}
\end{figure}

Figures \ref{fig:markov_xx} and \ref{fig:markov_xy} specify more closely the situations from the scenario 3 (with the TMA 1 not submitted, and TMA 1 passed, respectively). Colour tones of arrows differ (from white to red) depending on the percentage of students who moved in given direction. The more red the colour the bigger the percentage of students it represents. The rows represent the weeks in which VLE activity was measured. First row shows activity before the beginning of the course (Week 0), the other four rows capture the VLE activity in week 1, 2, 3 and 4 respectively. The columns represent the intervals of VLE activity - different colour of the node mean different interval. First column is zero VLE activity, while in other columns the activity is divided into intervals with the cut points in multiples of 30.

\begin{figure}
  \centering
  \includegraphics[width=\columnwidth]{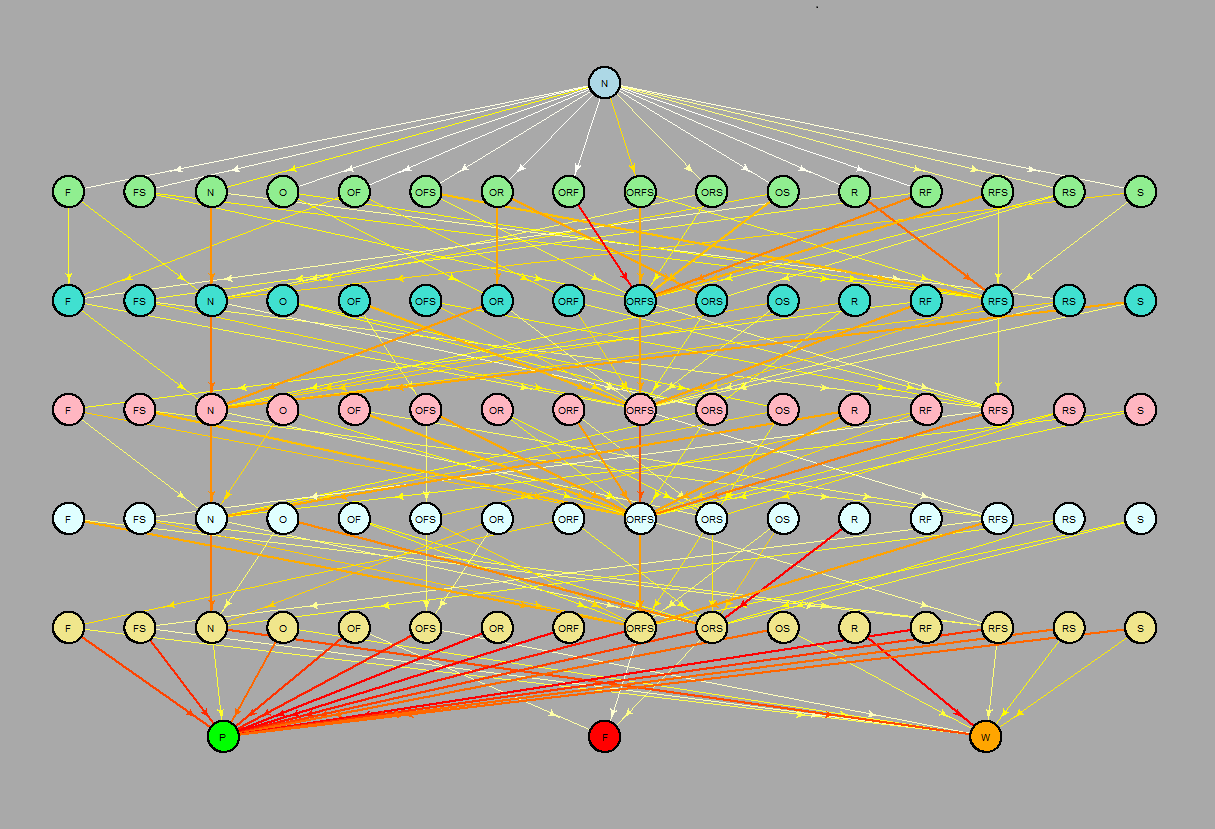}
  \caption{Markov chain for the various activity types combinations in weeks 0-5.}
  \label{fig:markov_activity_type}
\end{figure}

This type of analysis enables us to look for specific patterns in students' behaviour. In the similar way we can analyse the different types of VLE activities as shown in Figure \ref{fig:markov_activity_type}. In this case the nodes represent not the intensity of student activity, but capture the interest of student in specific content type. And (unlike the previous two figures) this directed acyclic graph as a whole depicts the Markov chain \cite{norris1998markov}.

\section{Conclusion}

In this paper we have examined two methods for analysing activity of students in the online virtual learning environment before the first tutor marked assignment -- GUHA and Markov chain-based graphical models. Both methods provide useful insights into the students' behaviour during their studies. The benefit of the latter lies mostly in its graphical output, which might be easier to interpret and could potentially provide support in planning interventions, and in the possibility to model time dependencies of the student activities. We believe that the understanding of the student behavioural patterns will also be useful for building better predictive models of student performance.

\bibliographystyle{acl}
\balance
\bibliography{bibliography}

\begin{thebibliography}{}

\bibitem[\protect\citename{Bishop and Nasrabadi},2006]{bishop2006pattern}
Christopher~M Bishop and Nasser~M Nasrabadi.
\newblock 2006.
\newblock {\em Pattern recognition and machine learning}, volume~1.
\newblock springer New York.

\bibitem[\protect\citename{Cormier},2008]{Cormier:2008:MOOC}
Dave Cormier.
\newblock 2008.
\newblock The {CCK08 MOOC}--connectivism course, 1/4 way.
\newblock {\em Dave's Educational Blog}, 2.

\bibitem[\protect\citename{H{\'a}jek \bgroup et al.\egroup
  },1966]{Hajek:1966:GUHA}
Petr H{\'a}jek, I.~Havel, and Michal Chytil.
\newblock 1966.
\newblock The guha method of automatic hypotheses determination.
\newblock {\em Computing}, 1(4):293--308.

\bibitem[\protect\citename{Koller \bgroup et al.\egroup
  },2013]{Koller:2013:MOOC_Retention}
Daphne Koller, Andrew Ng, Chuong Do, and Zhenghao Chen.
\newblock 2013.
\newblock Retention and intention in massive open online courses: In depth.
\newblock {\em EDUCAUSE}, June.

\bibitem[\protect\citename{Norris},1998]{norris1998markov}
James~R Norris.
\newblock 1998.
\newblock {\em Markov chains}.
\newblock Number 2008 in Cambridge series in statistical and probabilistic
  mathematics. Cambridge university press.

\bibitem[\protect\citename{Rauch and Simunek},2001]{Rauch:2001:4ft}
Jan Rauch and Milan Simunek.
\newblock 2001.
\newblock Mining for association rules by 4ft-miner.
\newblock In {\em INAP}, pages 285--295.

\bibitem[\protect\citename{Wolff and Zdrahal},2012]{Wolff:2012}
Annika Wolff and Zdenek Zdrahal.
\newblock 2012.
\newblock Improving retention by identifying and supporting "at-risk" students.
\newblock {\em EDUCAUSE Review Online}, July/Summer.

\bibitem[\protect\citename{Wolff \bgroup et al.\egroup },2013a]{Wolff:2014}
Annika Wolff, Zdenek Zdrahal, Drahomira Herrmannova, and Petr Knoth.
\newblock 2013a.
\newblock Predicting student performance from combined data sources.
\newblock In Alejandro Pe{\~n}a-Ayala, editor, {\em Educational Data Mining:
  Applications and Trends}, number 524 in Studies in Computational
  Intelligence, pages 175--202. Springer International Publishing, Cham.

\bibitem[\protect\citename{Wolff \bgroup et al.\egroup },2013b]{Wolff:2013}
Annika Wolff, Zdenek Zdrahal, Andriy Nikolov, and Michal Pantucek.
\newblock 2013b.
\newblock Improving retention: predicting at-risk students by analysing
  clicking behaviour in a virtual learning environment.
\newblock In {\em Third Conference on Learning Analytics and Knowledge (LAK
  2013)}.
\newblock ISBN 978-1-4503-1785-6.

\bibitem[\protect\citename{Wong and Chiu},1987]{Wong:1987:discretization}
Andrew K.~C. Wong and David K.~Y. Chiu.
\newblock 1987.
\newblock Synthesizing statistical knowledge from incomplete mixed-mode data.
\newblock {\em IEEE Trans. Pattern Anal. Mach. Intell.}, 9(6):796--805, June.

\end{thebibliography}

\end{document}